\documentclass[12pt]{iopart}

\newcommand{\Ih}{\mathrm{I\!h}}
\newcommand{\ri}{\mathrm{i}}
\newcommand{\rd}{\mathrm{d}}
\newcommand{\IR}{\mathbb{R}}

\usepackage{graphicx}
\usepackage{iopams}

\begin{document}

\title[Cylindrical equilibrium shapes of fluid membranes] 
{Cylindrical equilibrium shapes of fluid membranes} 
\author{V M Vassilev $^1$, P A Djondjorov $^1$ \MakeLowercase{and} I M Mladenov $^2$}
\address{$^1$ Institute of Mechanics, Bulgarian Academy of Sciences\\
Acad. G. Bonchev St., Block 4, 1113 Sofia, Bulgaria}
\address{$^2$ Institute of Biophysics, Bulgarian Academy of Sciences\\
Acad. G. Bonchev St., Block 21, 1113 Sofia, Bulgaria}
\ead{vasilvas@imbm.bas.bg}

\begin{abstract}
Within the framework of the well-known curvature models, a fluid lipid bilayer membrane is regarded as a surface embedded in the three-dimensional Euclidean space whose equilibrium shapes are described in terms of its mean and Gaussian curvatures by the so-called membrane shape equation.
In the present paper, all solutions to this equation determining cylindrical membrane shapes are found and presented, together with the expressions for the corresponding position vectors, in explicit analytic form. The necessary and sufficient conditions for such a surface to be closed are derived and several sufficient conditions for its directrix to be simple or self-intersecting are given.
\end{abstract}



\section{Introduction}

By a fluid membrane in this paper we assume a membrane formed in aqueous solution by a bilayer of lipid molecules, which are in a fluid
state, i.e., the molecules can move freely within the monolayer
they belong to. The structure of the bilayer is such that the hydrophobic tails of the lipid molecules situated in different monolayers face one another to form a semipermeable core, while their hydrophilic heads face the aqueous solutions on either side of the membrane. It is known that the lipid bilayer is the main structural component of all biological membranes, the closed lipid bilayer membranes (vesicles) being thought of as the simplest model systems for studying basic physical properties of the more complex biological cells.

The foundation of the current theoretical understanding of the
vesicle shapes (see, e.g. \cite{LSh, ZCan, Seifert}) can be
traced more than thirty years back to the works by Canham
\cite{Canham} and  Helfrich \cite{Helf},
in which the so-called curvature models have been introduced. In
these models, the vesicle's membrane is regarded as a
two-dimensional surface $\mathcal{S}$ embedded in the
three-dimensional Euclidean space $\mathrm{I\!R}^{3}$ and assumed
to exhibit purely elastic behaviour described by its mean $H$ and
Gaussian $K$ curvatures and two material constants associated with
the bending rigidity of the membrane.

In the model proposed by Helfrich \cite{Helf}, currently referred
to as the spontaneous-curvature model, the equilibrium shapes of
the vesicles are determined by the extremals of the curvature
(shape) energy
\begin{equation*}
\mathcal{F}_{c}=\frac{k_{c}}{2}\int_{\mathcal{S}}\left(
2H+\Ih\right) ^{2}{
\mathrm{d}}A+k_G\int_{\mathcal{S}}K{\mathrm{d}}A
\label{ShE}
\end{equation*}
under the constraints of fixed enclosed volume $V$ and total
membrane area $A$. This scheme yields the functional
\begin{equation}
\mathcal{F}=\frac{k_{c}}{2}\int_{\mathcal{S}}\left( 2H+\Ih\right) ^{2}{%
\mathrm{d}}A+k_G\int_{\mathcal{S}}K{\mathrm{d}}A+\lambda \int_{\mathcal{S%
}}{\mathrm{d}}A+p\int {\mathrm{d}}V.
\label{HFFunc}
\end{equation}
Here $k_{c}$ and $k_G$ are real constants representing the bending
and Gaussian rigidity of the membrane, $\Ih$ is the spontaneous curvature (a constant introduced by Helfrich to reflect the
asymmetry of the membrane or its environment), $p$ and $\lambda$
are the Lagrange multipliers (another two real constants)
corresponding to the constraints of enclosed volume and total
area, respectively, whose physical meaning is as follows: $p$
represents the pressure difference between the outer and inner
sides of the membrane, while $\lambda$ is interpreted as a tensile
stress
or a chemical potential (see, e.g., \cite{Brown}).
The Euler-Lagrange equation corresponding to the 
functional (\ref{HFFunc}) reads
\begin{equation}
2k_{c}\Delta H+k_{c}\left( 2H+\Ih\right) (2H^{2}-\Ih
H-2K)-2\lambda H+p=0, \label{MShEq}
\end{equation}
where $\Delta $ is the Laplace-Beltrami operator on the surface
$\mathcal{S}$. This equation, derived in
\cite{ ZcHelf2},
is often referred to as the membrane shape equation.
It is worth noting that the second term in the curvature energy $\mathcal{F}_{c}$ does not effect equation (\ref{MShEq}) since its contribution to the overall Lagrangian density is a total divergence as follows from the Liouville's form of Gauss's Theorema Egregium (see, e.g., \cite{Coxeter}). Actually, for closed membranes without edges the integral over the Gaussian curvature $K$ is a topological invariant by virtue of the Gauss-Bonnet theorem and therefore it may be disregarded until the topology of the membrane remains unchanged. This term, however, plays an important role when topological phase transitions are considered (see, e.g., \cite{PhT1, PhT2}) as well as in the theory of fluid membranes with free edges (cf. \cite{ FE2, FE1, FE3}).

Later on, another two curvature models have been developed. The first
of them is the so-called bilayer-couple model suggested by Svetina
and \v{Z}ek\v{s} in \cite{SZ} on the ground of the bilayer-couple
hypothesis \cite{SS} and the related work \cite{
SZ1}. The second one is known as the area-difference-elasticity
model \cite{BSZW, Miao, WHH}. For the purposes of the present
paper, however, it is important to underline that all the
curvature models mentioned above lead to the same set of
stationary shapes, determined locally by equation (\ref{MShEq}),
since they differ only in global energy terms (see \cite{LSh,
Miao, SZ}). Of course, the constants involved in this equation have
different meaning in different models.

For more than three decades, the study of the equilibrium shapes
of the vesicles has attracted much attention nevertheless to the best of our knowledge only a few analytic solutions to equation (\ref{MShEq}) have been reported.
These are the solutions determining: spheres and
circular cylinders \cite{ZcHelf2}, Clifford tori \cite{CT}
and toroidal shapes (cf. \cite{LSh}, Ch.~8, Sect.~5),
circular biconcave discoids \cite{BCC}, Delaunay surfaces
\cite{mla,Naito}, nodoidlike and unduloidlike shapes \cite{Naito},
several types of surfaces with constant squared mean curvature density and Willmore surfaces (cf. \cite{Konopelchenko, VM, Willmore})
as well as cylindrical surfaces corresponding to $p \neq 0$ \cite{VDM, VDM1} and $p=0$.
The latter case is exceptional since it coincides with the prominent Euler's elastica whose typical equilibrium shapes are known for a long time \cite{Love}. This problem has been studied for more than three centuries in various contexts, see, e.g., the recent papers \cite{bru, IS, LS, Linner, ZOY}.

The main goal of the present paper is to present in analytic form all
solutions to the membrane shape equation (\ref{MShEq}) determining cylindrical surfaces as well as to provide explicit expressions for
the corresponding position vectors. In a sense, it might be thought of as a completion, from analytic point of view, of the works by Arreaga {\it et al.} \cite{ACCG} and Capovilla {\it et al.} \cite{CCG} where a purely geometric construction of the equilibrium shapes of closed planar loops subject to the constraints of fixed length and enclosed area is presented.

\section{Cylindrical equilibrium shapes}

For cylindrical surfaces in $\mathrm{I\!R}^{3}$ whose directrices are plane curves $\Gamma $ of curvature $\kappa(s)$ parametrized by their arclength $s$,
the corresponding generatrices being perpendicular to the plane the directrices $\Gamma $ lie in (see figure~1),
the general shape equation (\ref{MShEq}) simplifies and reduces to the single ordinary differential equation
\begin{equation}
2\frac{{\mathrm{d}}^{2}\kappa\left( s\right)}{{\mathrm{d}}s^{2}}+\kappa\, ^{3} \left( s\right)-\mu \kappa\left( s\right)-\sigma =0, \label{HEq}
\end{equation}
where
\[
\mu =\Ih^{2}+\frac{2\lambda }{k_{c}},
\qquad
\sigma = -\frac{2p}{k_{c}} \,\cdot
\]
Indeed, using the standard formulas from the textbooks on classical differential geometry (see, e.g., \cite{Coxeter, Oprea}) one can easily find that for a cylindrical surface parametrized in the above way $H=(1/2)\kappa(s)$ and
$\Delta H = (1/2)\mathrm{d}^{2}\kappa(s)/\mathrm{d}s^{2}$
in addition to the relation $K=0$. 
Substituting the latter expressions in equation (\ref{MShEq}) one gets immediately to equation (\ref{HEq}).
\begin{figure}[h]
\centering
\includegraphics[width=4.2in]{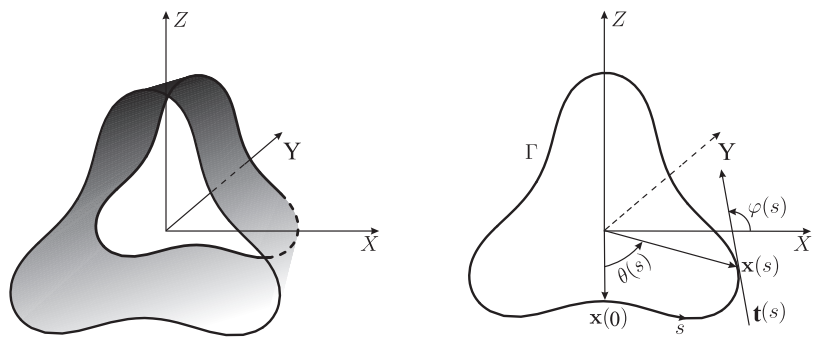}
\caption{A slice of the infinite generalized cylinder (left) and
its intersection with the plane $Y\equiv0$ (right). Here, ${\bf t}(s)$, $\varphi(s)$ and $\theta(s)$ are the tangent vector, slope angle and the angle between the position vectors $\mathbf{x}(0)$ and $\mathbf{x}(s)$, respectively.}
\end{figure}

In what follows, we are interested in real-valued solutions $\kappa(s) \neq \mathrm{const}$ of equation (\ref{HEq}) possessing smooth derivatives. Once such a solution is known, it is possible to recover the embedding
$\mathbf{x}\left( s\right)=\left( x(s),z(s) \right)\in\mathrm{I\!R}^{2}$
of the corresponding directrix $\Gamma $ in the plane
$\mathrm{I\!R}^{2}$ (up to a rigid motion) in the standard
manner. First, recall that the unit tangent $\mathbf{t}\left(
s\right)=(\mathrm{d}x\left( s\right) /
\mathrm{d}s,\mathrm{d}z\left( s\right) / \mathrm{d}s)$ and normal
$\mathbf{n}\left( s\right)=(-\mathrm{d}z\left( s\right) /
\mathrm{d}s,\mathrm{d}x\left( s\right) / \mathrm{d}s)$ vectors to
the curve $\Gamma $ are related to the curvature $\kappa(s)$ through Frenet-Serret formulas \cite{Coxeter, Oprea}
\begin{equation}
\frac{{\mathrm{d}}\mathbf{t}\left( s\right) }{{\mathrm{d}}s}= \kappa%
\left( s\right) \mathbf{n}\left( s\right) ,\qquad \frac{{\mathrm{d}}\mathbf{n}%
\left( s\right) }{{\mathrm{d}}s}= - \kappa\left( s\right)
\mathbf{t}\left( s\right) .  \label{Frenet}
\end{equation}
Consequently, in terms of the slope angle $\varphi\left( s\right)$
of the curve $\Gamma$ (see figure~1), one has
\begin{equation}
\kappa\left( s\right) = \frac{\mathrm{d}\varphi \left( s\right)
}{\mathrm{d}s},\qquad \frac{\mathrm{d}x\left(
s\right) }{\mathrm{d}s}=\cos \left( \varphi \left( s\right) \right),
\qquad \frac{\mathrm{d}z\left( s\right) }{\mathrm{d}s}=\sin \left( \varphi
\left( s\right) \right) \label{Phy}
\end{equation}
and hence, the parametric equations of the curve $\Gamma $ can be expressed by quadratures
\begin{equation}
x\left( s\right) =\int\cos \left( \varphi \left( s\right) \right)
\mathrm{d}s,\qquad z\left( s\right) =\int\sin \left( \varphi \left(s\right) \right) \mathrm{d}s, \label{xz}
\end{equation}
where
\begin{equation}
\varphi \left( s\right) =\int\kappa\left( s\right) \mathrm{d}s.
\label{angle}
\end{equation}
Thus, the first problem to solve on the way to determining the cylindrical equilibrium shapes of the fluid membranes is to find the solutions of equation (\ref{HEq}) in analytic form.

Fortunately, equation (\ref{HEq}) is integrable by quadrature since it falls into the class of equations describing conservative systems with one degree of freedom \cite{Arnold}. Indeed, (\ref{HEq}) can be regarded as the equation of motion of a fictitious particle of unit mass  whose kinetic, potential and total energies are
\[
T=\frac{1}{2}\left( \frac{{\mathrm{d}}\kappa}{{\mathrm{d}}s}\right)^{2},
\qquad
U(\kappa)=\frac{1}{8}\kappa^{4}-\frac{1}{4}\mu
\kappa^{2}-\frac{1}{2}\sigma \kappa,
\qquad
\mathcal{E}=T+U,
\]
respectively. In this setting, $\kappa$ is interpreted as the displacement of the particle while $s$ plays the role of the time.
The total energy $\mathcal{E}$ of this system is conserved and hence
\begin{equation}
\left( \frac{{\mathrm{d}}\kappa}{{\mathrm{d}}s}\right) ^{2}=P(\kappa),
\qquad P(\kappa) =2E-\frac{1}{4}\kappa^{4}+\frac{1}{2}\mu
\kappa^{2}+\sigma \kappa
\label{CCCL}
\end{equation}
holds on each continuous solution of equation (\ref{HEq}), $E$ being the value of its total energy. Therefore, the solution of equation (\ref{HEq}) can be reduced to the quadrature
\begin{equation}
s=\int \frac{{\mathrm{d}}\kappa }{\sqrt{2(E-U(\kappa)) }}=
\int \frac{{\mathrm{d}}\kappa }{\sqrt{P(\kappa) }}
\label{quadr}
\end{equation}
up to a shift of the independent variable $s$ and change of its sign to the opposite one. Note that equation (\ref{HEq}) and its first integral (\ref{CCCL}) are invariant under the aforementioned transformations of the variable $s$.
Note also that for each solution $\kappa=\kappa(s)$ of equation (\ref{HEq}) relation (\ref{CCCL}) implies the existence of a certain value of the variable $s$ at which ${\mathrm{d}}\kappa /{\mathrm{d}}s=0$
(this matter will be clarified in detail at the beginning of the next Section).
Without loss of generality, this value may be chosen to be zero due to the translational invariance of equation (\ref{HEq}). So, hereafter we will always chose ${\mathrm{d}}\kappa /{\mathrm{d}}s=0$ at $s=0$.

Moreover, the specific differential structure of equation (\ref{HEq}) allows the integration in expressions (\ref{xz}) to be avoided when $\sigma \neq0$. Below, one can find a simple alternative derivation of this remarkable integrability property, first established in \cite{ACCG} (see also \cite{CCG, GV}). Actually, a direct computation shows that the following identity holds
\begin{eqnarray}
\fl \eqalign{\left (2\frac{{\mathrm{d}}^{2}
\kappa (s)}{{\mathrm{d}}s^{2}}+
\kappa\,^{3}(s)-\mu \kappa (s)-\sigma \right ) \mathbf{t}(s)+ 2\frac{{\mathrm{d}}\kappa(s) }{{\mathrm{d}}s}
\left( \frac{{\mathrm{d}}\mathbf{t}(s)}{{\mathrm{d}}s}-
\kappa(s) \mathbf{n}(s)\right) \\
-( \kappa\,^{2} (s)-\mu)
\left( \frac{{\mathrm{d}}\mathbf{n}(s)}{{\mathrm{d}}s}+\kappa(s)
 \mathbf{t}(s)\right)=
\frac{{\mathrm{d}}}{{\mathrm{d}}s}\left( 2\frac{{\mathrm{d}}\kappa(s)}
{{\mathrm{d}}s}\mathbf{t}(s)-
(\kappa\, ^{2}(s)-\mu ) \mathbf{n}(s)-\sigma \mathbf{x}(s)
\right) }
\label{identity}
\end{eqnarray}
and hence, taking into account Frenet-Serret formulas (\ref{Frenet}), one can represent the position vector $\mathbf{x}\left( s\right)$
of a plane curve of curvature $\kappa \left( s\right) $ in the form
\begin{equation*}
\mathbf{x}\left( s\right) =\frac{2}{\sigma }\frac{{\mathrm{d}}\kappa%
\left( s\right) }{{\mathrm{d}}s}\mathbf{t}\left( s\right)-\frac{1}{\sigma }
( \kappa \,^{2}( s) -\mu ) \mathbf{n}\left(s\right) +\mathbf{C},
\label{embedding}
\end{equation*}
where $\mathbf{C}$ is a constant vector, provided that
$\kappa\left( s\right)$ is a solution of equation (\ref{HEq}) with
$\sigma \neq 0$.
Then, translating the origin so that
$\mathbf{x} \cdot \mathbf{t} =0$ and
$\mathbf{x} \cdot \mathbf{n} =-1/\sigma \left( \kappa ^{2}-\mu \right)$
when ${\mathrm{d}}\kappa /{\mathrm{d}}s=0$, which is always possible, one gets $\mathbf{C}=\mathbf{0}$ and obtains, taking into account the definitions of the tangent and normal vectors as well as the second and the third of the relations (\ref{Phy}), the following expressions for the components of the position vector in therms of the curvature $\kappa(s)$ and slope angle $\varphi(s)$
\begin{eqnarray}
\eqalign{x(s) =\frac{2}{\sigma}\frac{\mathrm{d}\kappa(s)}{\mathrm{d}s}
\cos \varphi (s) + \frac{1}{\sigma }( \kappa\, ^{2}(s)-\mu )
\sin \varphi (s),
\label{Sol1} \\
z(s) =\frac{2}{\sigma }\frac{\mathrm{d}\kappa(s) }{\mathrm{d}s}
\sin \varphi (s) - \frac{1}{\sigma } ( \kappa \,^{2}( s)-\mu )
\cos \varphi (s). }
\end{eqnarray}
Note, however, that the slope angle $\varphi(s)$ still remains determined implicitly, via an integration (\ref{angle}) of the curvature $\kappa(s)$, which, so far, cannot be accomplished. Note also that formulas
(\ref{CCCL}) and (\ref{Sol1}) lead to the remarkable relation
\begin{equation}
r^2(s)= \frac{8E+\mu ^{2}}{\sigma^{2}} +
\frac{4 \kappa(s)}{\sigma}
\label{xx}
\end{equation}
for the magnitude $r(s)=\sqrt{x^2(s)+z^2(s)}$ of the position vector $\mathbf{x}(s)$ found in \cite{ACCG, CCG}.

For $\sigma=0$, the situation is quite similar. In this case, identity (\ref{identity}) implies
\begin{equation}
\frac{{\mathrm{d}}\kappa(s) }{{\mathrm{d}}s}\mathbf{t}(s)-
\frac{1}{2} ( \kappa^{2}(s)-\mu ) \mathbf{n}(s)=\mathbf{C_0},
\label{is0}
\end{equation}
where $\mathbf{C_0} $ is a constant vector, provided that
$\kappa\left( s\right)$ is a solution of equation (\ref{HEq}) with
$\sigma= 0$ and relations (\ref{Frenet}) hold. Now, choosing
$\varphi={\mathrm{d}}\kappa /{\mathrm{d}}s=0$ at $s=0$ and taking into account relation (\ref{CCCL}), the definitions of the tangent and normal vectors as well as the second and the third of the relations (\ref{Phy}), one can first see that $\mathbf{C_0}=( 0, -( \kappa ^{2}(0)-\mu )/2)\neq \mathbf{0}$ and then rewrite equality (\ref{is0}) in the form
\begin{equation}
\cos \varphi \left( s\right) =\frac{\kappa ^{2}\left( s\right) -\mu }{\kappa ^{2}\left( 0\right)-\mu },\qquad
\sin \varphi \left( s\right) =-\frac{2}{\kappa ^{2}\left( 0\right)-\mu} \frac{{\mathrm{d}}\kappa \left( s\right) }{{\mathrm{d}}s} \,\cdot
\label{cs}
\end{equation}
Consequently, formulas (\ref{xz}) and (\ref{cs}) imply
\begin{equation}
x(s)=\frac{1}{\kappa ^{2}\left( 0\right)-\mu}
\int \kappa ^{2} \left(s\right){\mathrm{d}}s
-\frac{\mu s}{\kappa ^{2}\left( 0\right)-\mu}
, \quad
z\left( s\right) =-\frac {2 \kappa \left( s\right)}
{\kappa ^{2}\left( 0\right)-\mu }\,\cdot
\label{xz0}
\end{equation}
Again, one integration remains to be done. This time, the square of the curvature $\kappa(s)$ has to be integrated.

Before proceeding with the derivation of the solutions of equation (\ref{HEq}) using the quadrature (\ref{quadr}), it should be remarked that this equation have been regarded in a number of papers, see \cite{Fukumoto, TO, Watanabe, ZTHL}, that have not been mentioned yet since they do not concern directly the problem considered here. Closest to the subject of the present paper are \cite{ACCG, CCG} where equation (\ref{HEq}) is introduced with the aim to study the equilibria of an elastic loop in the plane subject to the constraints of fixed length and enclosed area. In the three dimensional case considered here, each such loop determines a directrix $\Gamma $ of a cylindrical surface whose mean curvature satisfies the membrane shape equation.
In the foregoing two papers, the authors have succeeded in obtaining a purely
geometric construction for determination of the curvature of the
loop passing through a given point of the plane without using
explicit expressions for the solutions of equation (\ref{HEq}).
Nevertheless, in our opinion, the knowledge of the solutions to this equation in analytic form is of considerable interest for the further exploration of the cylindrical equilibrium shapes of the fluid membranes.

\section{Explicit analytic solutions}

First, it should be remarked that the quadrature (\ref{quadr}) can be easily expressed in terms of elliptic integrals
\cite{elliptic}
or elementary functions by means of the roots of the polynomial $P(\kappa)$ provided that the following observations are taken into account.

Given a solution $\kappa(s)$ of an equation of form (\ref{HEq}) with coefficients $\mu$ and $\sigma$, let $E$ be the value of its total energy.
Then, bearing in mind that $\mu$, $\sigma$ and $E$ are real numbers, it is clear that the corresponding polynomial $P(\kappa)$ has at least two different real roots, otherwise the function $P(\kappa(s))$ could not take different nonnegative values, as required by relation (\ref{CCCL}) and the assumptions concerning the type of solutions considered, since the coefficient at the highest power $\kappa^4$ of this polynomial is negative.
Thus, in general, only two alternative possibilities have to be considered, namely:
(I)~the polynomial $P(\kappa)$ has two simple real roots $\alpha,\beta \in \IR$, $\alpha <\beta$, and a pair of complex conjugate roots
$\gamma ,\delta \in \mathbb{C}$, $\delta =\bar{\gamma}$;
(II)~the polynomial $P(\kappa)$ has four simple real roots  $\alpha<\beta<\gamma <\delta \in \IR$.
In the first case, the polynomial $P(\kappa)$ is nonnegative in the interval $\alpha \leq \kappa \leq \beta $, while in the second one, it is nonnegative in the intervals
$\alpha \leq \kappa \leq \beta $ and $\gamma \leq \kappa \leq \delta $.

It should be noted also that the roots $\alpha ,\beta ,\gamma $ and
$\delta $ of the polynomial $P(\kappa)$ can be expressed explicitly
through its coefficients $\mu$, $\sigma$ and $E$ and vice versa. Indeed, after some standard algebraic manipulations
(cf. \cite{Korn}, Sec.~1.8), one can find the following expressions for the roots of the polynomial $P(\kappa)$
\begin{eqnarray*}
-\sqrt{\frac{\omega }{2}}-\sqrt{\mu -
\sigma \sqrt{\frac{2}{\omega }}-\frac{\omega }{2}}\,, \qquad
-\sqrt{\frac{\omega }{2}}+\sqrt{\mu -
\sigma \sqrt{\frac{2}{\omega }}- \frac{\omega }{2}}\,, \\[2mm]
\quad \sqrt{\frac{\omega }{2}}-\sqrt{\mu +
\sigma \sqrt{\frac{2}{\omega }}-\frac{\omega }{2}} \,, \qquad 
\phantom{-}\sqrt{\frac{\omega }{2}}+\sqrt{\mu +
\sigma \sqrt{\frac{2}{\omega }}-\frac{\omega }{2}}\,,
\end{eqnarray*}where
\begin{eqnarray*}
\omega &=&\frac{\left[ \mu +\sqrt[3]{3\left( 3^{2}\sigma ^{2}+\sqrt{\chi }\right) -\mu \left(\mu^{2}+2^{3}3^{2}E\right) }\, \right] ^{2}-2^{3}3E}{3\sqrt[3]
{3\left( 3^{2}\sigma ^{2}+\sqrt{\chi }\right) -\mu \left(
\mu^{2}+2^{3}3^{2}E\right) }}\,, \\[2mm]
\chi &=&3\left \{ 2^{3}E\left[ \left( \mu ^{2}+8E\right)
^{2}-3^{2}2\mu \sigma ^{2}\right] -\sigma ^{2}\left( 2\mu
^{3}-3^{3}\sigma ^{2}\right) \right \}.
\end{eqnarray*}
Then, one can denote properly each of the above expressions in
accordance with the notation introduced in the cases (I) and (II), respectively. Simultaneously, by Vieta's formulas one obtains
\begin{equation}
\alpha +\beta +\gamma +\delta = 0
\label{RCond}
\end{equation}
due to the absence of a term with $\kappa^3$ in the polynomial $P(\kappa)$,
and consequently
\begin{eqnarray}
\mu &=& \frac{1}{2}\left(\alpha^2+\beta^2+\gamma^2+ \alpha\beta+\alpha \gamma+\beta \gamma\right), \label{Coeffs1}\\
\sigma &=& -\frac{1}{4}\left( \alpha+ \beta\right)
\left( \alpha + \gamma \right) \left( \beta + \gamma \right), \label{Coeffs2} \\
E &=& \frac{1}{8} \alpha \beta \gamma \left( \alpha + \beta\ + \gamma \right).
\label{Coeffs3}
\end{eqnarray}

Now, we are in a position to express the arclength as a function of the curvature representing the quadrature (\ref{quadr}) via elliptic integrals 
or elementary functions in each of the particular cases (I) and (II). Instead of this, however, taking the corresponding inverse functions we prefer to give directly the curvature as a function of the arclength in terms of the roots of the polynomial $P(\kappa)$.
Moreover, solving the integral (\ref{angle}), we give explicit formulas for the corresponding slope angles as well.
The explicit analytic expressions for the solutions of equation (\ref{HEq}) and the corresponding slope angles can be found in Theorem~1, for the case (I), and in Theorem~2, for the case (II). 
\smallskip

\noindent \textbf{Theorem~1} \textit{Given $\mu$, $\sigma$ and $E$, let
the roots $\alpha ,\beta ,\gamma $ and $\delta $ of the respective polynomial $P(\kappa)$ are as in the case} (I), \textit{that is}
$\alpha <\beta \in \IR$,
$\gamma ,\delta \in \mathbb{C}$, $\delta =\bar{\gamma}$,
\textit{and let}
$\eta \equiv(\gamma -\bar{\gamma})/2\ri$.
\textit{Then, except for the cases in which}
$( 3\alpha +\beta)( \alpha +3\beta)=\eta=0$,
\textit{the function}
\begin{equation}
\kappa_1 \left( s\right) =\frac{\left( A\beta +B\alpha \right) - \left(
A\beta -B\alpha \right) \mathrm{cn}(us,k)}{\left( A+B\right)
- \left( A-B\right) \mathrm{cn}(us,k)} \label {FK3}
\end{equation}
\textit{of the real variable $s$, where}
\begin{eqnarray}
\hspace*{-1cm}
A=\sqrt{4\eta^{2}+\left( 3\alpha +\beta \right) ^{2}},\qquad
B=\sqrt{4\eta^{2}+\left( \alpha
+3\beta \right) ^{2}} \label {lab}, \qquad
u=\frac{1}{4}\sqrt{AB}\,, \\[2mm]
\hspace*{-1cm}k=\frac{1}{\sqrt{2}}\sqrt
{ 1-\frac{4\eta^{2}+\left( 3\alpha +\beta \right) \left(
\alpha +3\beta \right) }{ \sqrt{\left[ 4\eta^{2}+\left( 3\alpha +\beta \right) \left(
\alpha +3\beta \right)\right ]^2+16 \eta^2 (\beta - \alpha)^2}}\,,  }  \label {um2}
\end{eqnarray}
\textit{takes real values for each $s\in \IR$ and satisfies equation (\ref
{HEq}). This function is periodic if and only if $\eta \neq 0$ or $\eta =0$ and
$( 3\alpha +\beta)( \alpha +3\beta)>0$, its least period is
$T_1=\left( 4/u\right) \mathrm{K}\left( k\right) $, and for $\sigma \neq 0$ its indefinite integral $\varphi_1 \left( s \right)$ such
that $\varphi_1 \left( 0 \right)=0$ is}
\begin{eqnarray}
\hspace*{-1cm}\eqalign{\varphi_1 \left( s\right) =\frac{A\beta -B\alpha
}{A-B}s+\frac{\left( A+B\right) \left( \alpha -\beta \right)
}{2u\left( A-B\right) }\Pi \left
(-\frac{\left( A-B\right)^{2}}{4AB},
\mathrm{am}( us,k) ,k\right) \\
\hspace*{1.2cm} + \frac{\alpha -\beta }{2u\sqrt{k^2+\frac{\left( A-B\right) ^{2}}{4AB}}}%
\arctan \left( \sqrt{k^2+\frac{\left( A-B\right)
^{2}}{4AB}}\frac{\mathrm{sn}\left( us,k\right) }{\mathrm{dn}\left(
us,k\right) }\right)\cdot }
\label{Ang33}
\end{eqnarray}
\textit{In the cases in which}
$( 3\alpha +\beta)( \alpha +3\beta)=\eta=0$,
\textit{the function}
\begin{equation}
\kappa_2 \left( s\right) = \zeta - \frac{4 \zeta}{1+\zeta^{2}s^{2}}\,,
\label{H3}
\end{equation}
\textit{where $\zeta=\alpha$ when $3 \alpha + \beta=0$ and $\zeta=\beta$ when $\alpha +3\beta=0$, satisfies equation (\ref {HEq})
for each $s\in \IR$ and its indefinite integral
$\varphi_2 \left( s \right)$ such that $\varphi_2 \left( 0 \right)=0$ reads}
\begin{equation}
\varphi_2 \left( s\right) = \zeta s - 4 \arctan\left( \zeta s \right).
\label{AngleTripple}
\end{equation}
\smallskip

\noindent \textbf{Proof.}
Let us begin with the simpler case
$( 3\alpha +\beta)( \alpha +3\beta)=\eta=0$.
Here, on account of the definition of $\eta$ and relations
(\ref{RCond})--(\ref{Coeffs3}),
a straightforward computation shows that the function (\ref
{H3}) is the derivative of the function (\ref{AngleTripple}) and satisfies
equation (\ref {HEq}).
The relation $\varphi_2 \left( 0 \right)=0$ is obvious.

Next, let us exclude the cases in which
$( 3\alpha +\beta)( \alpha +3\beta)=\eta=0$. Now, the condition
$\alpha <\beta \in \IR$, the definition of $\eta$
and expressions (\ref {lab}) and (\ref{um2})
imply also that
$\eta, A, B, u, k\in \IR$, $AB \neq 0$, $u>0$ and $0 \leq k \leq 1$.
Hence, the function (\ref {FK3}) is real-valued when $s\in \IR$.
Substituting the function (\ref {FK3}) into equation (\ref {HEq})
and taking into account relations (\ref{RCond})--(\ref{Coeffs3}),
one can easily verify that the latter equation is satisfied.
When $\eta \neq 0$, expression (\ref{um2}) implies $0 < k < 1$ and therefore
the function (\ref {FK3}) is periodic because the
function $\mathrm{cn}\left( us,k\right)$ is periodic, with least period $T_1$.
When $\eta = 0$ but $( 3\alpha +\beta)( \alpha +3\beta)\neq0$
two alternative cases are to be considered, namely $( 3\alpha +\beta)( \alpha +3\beta) > 0$ and $( 3\alpha +\beta)( \alpha +3\beta) < 0$. In the first case, expression (\ref{um2}) leads to $k=0$, which means that $\mathrm{cn}(us,k)=\mathrm{cos}(us)$ and hence the function (\ref {FK3}) is periodic again. However, if $( 3\alpha +\beta)( \alpha +3\beta) < 0$, then expression (\ref{um2}) implies $k=1$ meaning that $\mathrm{cn}(us,k)=\mathrm{sech}(us)$, and hence the function (\ref {FK3}) is not periodic.
Thus, having considered all possible cases, we may conclude that
the function (\ref {FK3}) is periodic if and only if
$\eta \neq 0$ or $\eta =0$ and $( 3\alpha +\beta)( \alpha +3\beta)>0$.
Finally, differentiation of expressions (\ref {Ang33}) with respect to the variable $s$ yields (\ref {FK3}). The relation
$\varphi_1 \left( 0 \right)=0$ is obvious.
$\blacksquare $
\smallskip

\noindent \textbf{Theorem~2} \textit{Given $\mu$, $\sigma$ and $E$, let
the roots $\alpha ,\beta ,\gamma $ and $\delta $ of the respective polynomial $P(\kappa)$ are as in the case} (II), \textit{that is}
$\alpha <\beta <\gamma <\delta \in \IR$.
\textit{Consider the functions }
\begin{equation}
\kappa_3 \left( s\right)  =\delta -\frac{\left( \delta -\alpha \right)
\left( \delta -\beta \right) }{\left( \delta -\beta \right)
+\left( \beta -\alpha \right) \mathrm{sn}^{2}\left( us,k\right)}\,,
\label {FK1}
\end{equation}
\begin{equation}
\kappa_4 \left( s\right)  =\beta +\frac{\left( \gamma -\beta \right)
\left( \delta -\beta \right) }{\left( \delta -\beta \right)
-\left( \delta -\gamma \right) \mathrm{sn}^{2}\left( us,k\right)}\,,
\label {FK2}
\end{equation}
\textit{of the real variable $s$, where }
\begin{equation}
u=\frac{1}{4}\sqrt{\left( \gamma -\alpha \right) \left( \delta
-\beta \right) },\qquad k=\sqrt \frac{\left( \beta -\alpha \right)
\left( \delta -\gamma \right) }{\left( \gamma -\alpha \right)
\left( \delta -\beta \right) } \cdot  \label {um1}
\end{equation}
\textit{Then, both functions (\ref{FK1}) and (\ref{FK2})
take real values for each $s\in \IR$ and satisfy equation (\ref {HEq}), they are periodic with least period
$T_2=\left( 2/u\right)\mathrm{K}\left( k\right) $
and their indefinite integrals $\varphi_3 \left( s \right)$ and $\varphi_4 \left( s \right)$, respectively, such that
$\varphi_3 \left( 0 \right)=\varphi_4 \left( 0 \right)=0$ are}
\begin{equation}
\varphi_3 \left( s\right) =\delta s-\frac{\delta -\alpha }{u}\Pi
\left( \frac{\beta -\alpha }{\beta -\delta
},\mathrm{am}(us,k),k\right),  \label{Angle1}
\end{equation}
\begin{equation}
\varphi_4 \left( s\right) =\beta s-\frac{\beta -\gamma }{u}\Pi
\left( \frac{ \delta -\gamma }{\delta -\beta
},\mathrm{am}(us,k),k\right).  \label{Angle2}
\end{equation}
\smallskip

\noindent \textbf{Proof.} It is easy to see that the condition
$\alpha <\beta <\gamma <\delta \in \IR$ and expressions
(\ref {um1}) imply also that $u,k\in \IR$, $u>0$ and $0<k<1$. Therefore,
both functions (\ref {FK1}) and (\ref {FK2}) are real-valued when $s\in \IR$.
Substituting each of the above functions into equation (\ref {HEq})
and taking into account relations (\ref{RCond})--(\ref{Coeffs3}),
one can easily verify that they satisfy it.
Evidently, these functions are periodic due to the fact that
the function $\mathrm{sn} ^{2}( us,k)$ is periodic,
with least period $T_2=\left( 2/u\right) \mathrm{K}\left( k\right) $,
since $u>0$ and $0<k<1$.
Finally, differentiation of expressions (\ref{Angle1}) and (\ref{Angle2}) with respect to the variable $s$ yields (\ref{FK1}) and (\ref{FK2}),
respectively. The relations $\varphi_3 \left( 0 \right)=\varphi_4 \left( 0 \right)=0$ are obvious. $\blacksquare $
\smallskip

Suppose now that $\sigma =0$. Under this assumption, in the case (I), formulas (\ref{RCond}) and (\ref{Coeffs2}) imply $\beta =-\alpha>0 $.
Then, according to Theorem~1, $B=A=2\sqrt{\eta ^{2}+\alpha ^{2}}$ and so
\begin{equation}
\kappa _{1}\left( s\right) =\alpha \,\mathrm{cn}(us,k), \qquad
u=\frac{1}{2}\sqrt{\eta ^{2}+\alpha ^{2}},\qquad
k=\sqrt{\frac{\alpha ^{2}}{\eta^{2}+\alpha ^{2}}}\,,
\label{C1k}
\end{equation}
cf. formulas (\ref {FK3}) -- (\ref {um2}). Consequently
\begin{equation}
\int \kappa \,_{1}^{2}\left( s\right) \mathrm{d}s
=2\sqrt{\alpha ^{2}+\eta ^{2}}\, 
\mathrm{E}\left( \mathrm{am}(us,k),k\right)-\eta ^{2}s.
\label{IntC1}
\end{equation}
Note also that formulas (\ref{C1k}) and (\ref{Coeffs1}) imply
\begin{equation}
\kappa _{1}\left( 0\right)=\alpha, \qquad
\mu=\frac{1}{2}(\alpha ^{2}-\eta ^{2}).
\label{mu1}
\end{equation}
In the case (II), the assumption $\sigma =0$ and formulas (\ref{RCond}) and (\ref{Coeffs2}) imply $\delta =-\alpha>0 $, $\gamma =-\beta>0$.
Hence, according to Theorem~2, cf. formulas (\ref {FK1})--(\ref {um1}), we have 
\begin{equation*}
u=-\frac{1}{4}\left( \alpha +\beta \right)= \frac{1}{4}\left( \delta + \gamma \right),
\qquad
k=-\frac{\beta -\alpha }{\beta +\alpha }
=\frac{\delta -\gamma }{\delta +\gamma },
\end{equation*}
and
\begin{equation*}
\kappa _{3}\left( s\right) =\alpha \frac{1-k\,\mathrm{sn}^{2}\left( us,k\right)
}{1+k\,\mathrm{sn}^{2}\left( us,k\right)},\qquad
\kappa _{4}\left(s\right) =\gamma \frac{1+k\,\mathrm{sn}^{2}\left( us,k\right) }{1-k\,\mathrm{sn}^{2}\left( us,k\right) }\cdot
\end{equation*}
Then, using Gauss's transformation $\tilde{u}=u\left( 1+k\right)$,
$\tilde{k}=2\sqrt{k}/(k+1)$, we obtain
\begin{equation}
\kappa _{3}(s) =\alpha \, \mathrm{dn}( \tilde{u}s,\tilde{k}), \qquad
\tilde{u}=-\frac{1}{2}\alpha, \qquad
\tilde{k}=-\frac{1}{\alpha}\sqrt{\alpha^2-\beta^2},
\label{C2}
\end{equation}
\begin{equation}
\kappa _{4}(s) = \gamma \, \frac{1 }{\mathrm{dn}( \tilde{u}s,\tilde{k})}=
-\alpha \, \mathrm{dn}( \tilde{u}s+\mathrm{K}(\tilde{k}),\tilde{k})=
-\kappa _{3}(s+\tilde{u}^{-1} \mathrm{K}(\tilde{k}))\cdot
\label{C3}
\end{equation}
Now, observing relation (\ref{C3}), we arrive at the conclusion that the Euler's elastic curves of curvatures $\kappa _{3}(s)$ and $\kappa _{4}(s)$ determined by formulas (\ref{C2}) and (\ref{C3}), respectively, coincide up to a rigid motion in the plane $\IR^{2}$. Therefore, to complete the present task it suffices to consider only the curves of curvature given by formula (\ref{C2}). In this case, we have
\begin{equation}
\int \kappa \,_{3}^{2}\left( s\right) \mathrm{d}s=
- 2 \alpha \mathrm{E} ( \mathrm{am}(\tilde{u}s,\tilde{k}),\tilde{k}) 
\label{IntC2_1}
\end{equation}
and
\begin{equation}
\kappa _{3}\left( 0\right)=\alpha, \qquad
\mu=\frac{1}{2}(\alpha ^{2}+\beta ^{2}).
\label{mu2}
\end{equation}
due to the relations (\ref{C2}) and (\ref{Coeffs1}).

Thus, having obtained in explicit form the solutions of equation (\ref {HEq}), i.e., the curvatures, the corresponding slope angles and the integrals of the squared curvatures (in the cases in which $\sigma = 0$), we have completely determined in analytic form the corresponding directrices $\Gamma $ (up to a rigid motion in the plane $\IR^{2}$) through the parametric equations (\ref{Sol1})
when $\sigma \neq 0$ or (\ref{xz0}), with the supplementary relations (\ref{C1k})--(\ref{mu1}) or (\ref{C2}), (\ref{IntC2_1}) and (\ref{mu2}), when $\sigma = 0$.
Several examples of directrices of cylindrical equilibrium shapes corresponding to solutions to equation (\ref{HEq}) of form (\ref {FK3}), (\ref {FK1}) or (\ref {FK2}) are presented in figure~2 and figure~3 for $\sigma \neq 0$ and in figure~4 for $\sigma = 0$.

\begin{figure}[h]
\centerline{
\hspace*{6mm}
\includegraphics[height=3.0cm]{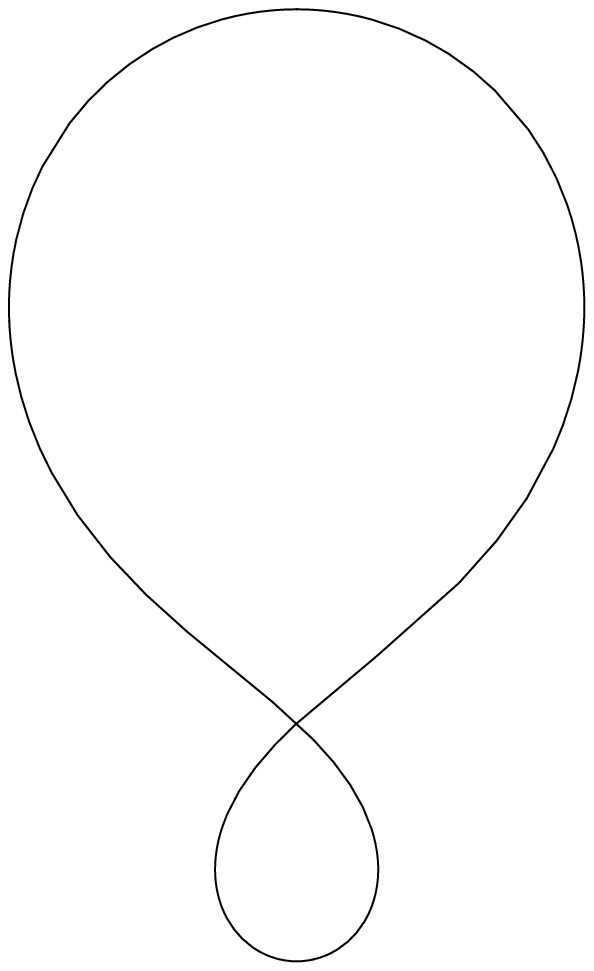}
a) \hspace*{5mm}
\includegraphics[width=3.0cm, angle=90]{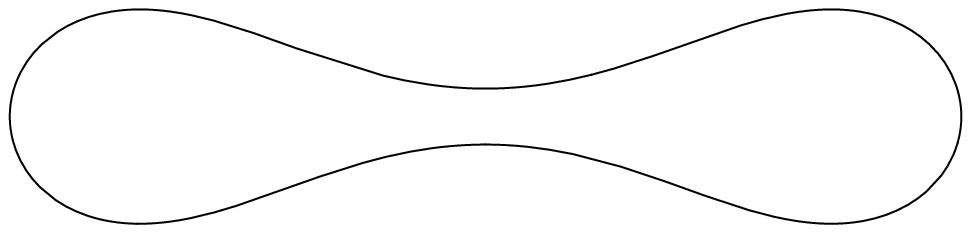}
b) \hspace*{5mm}
\includegraphics[height=3.0cm]{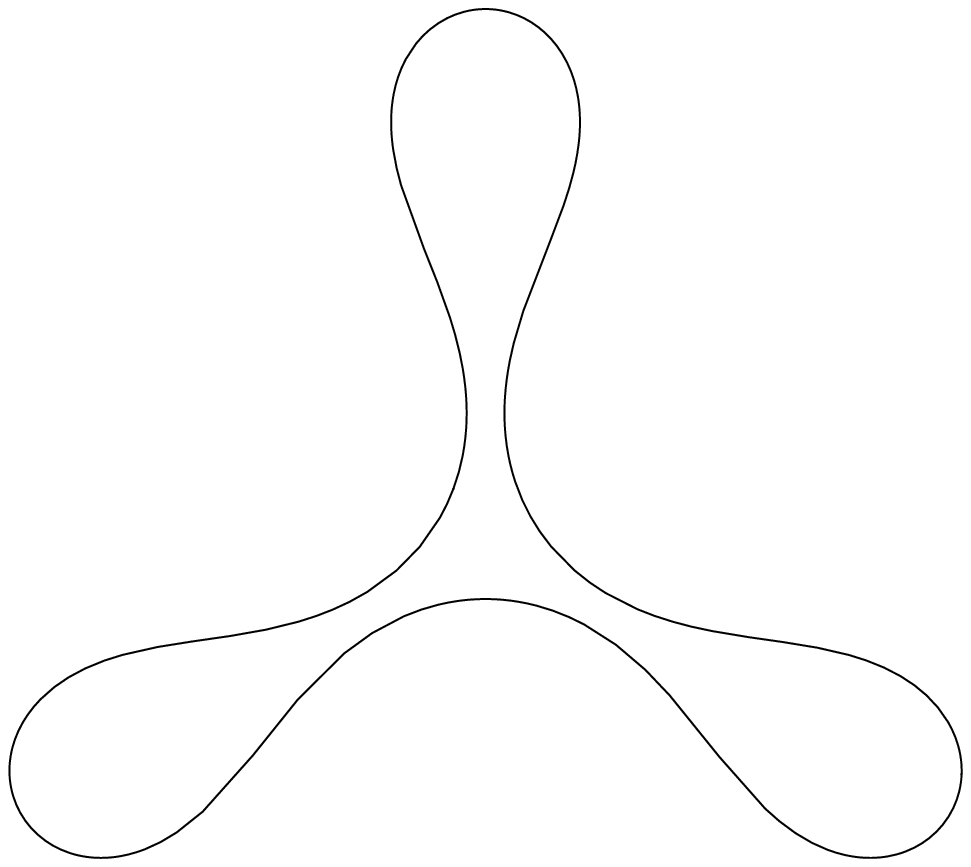}
c) \hspace*{4mm}
\includegraphics[height=3.0cm]{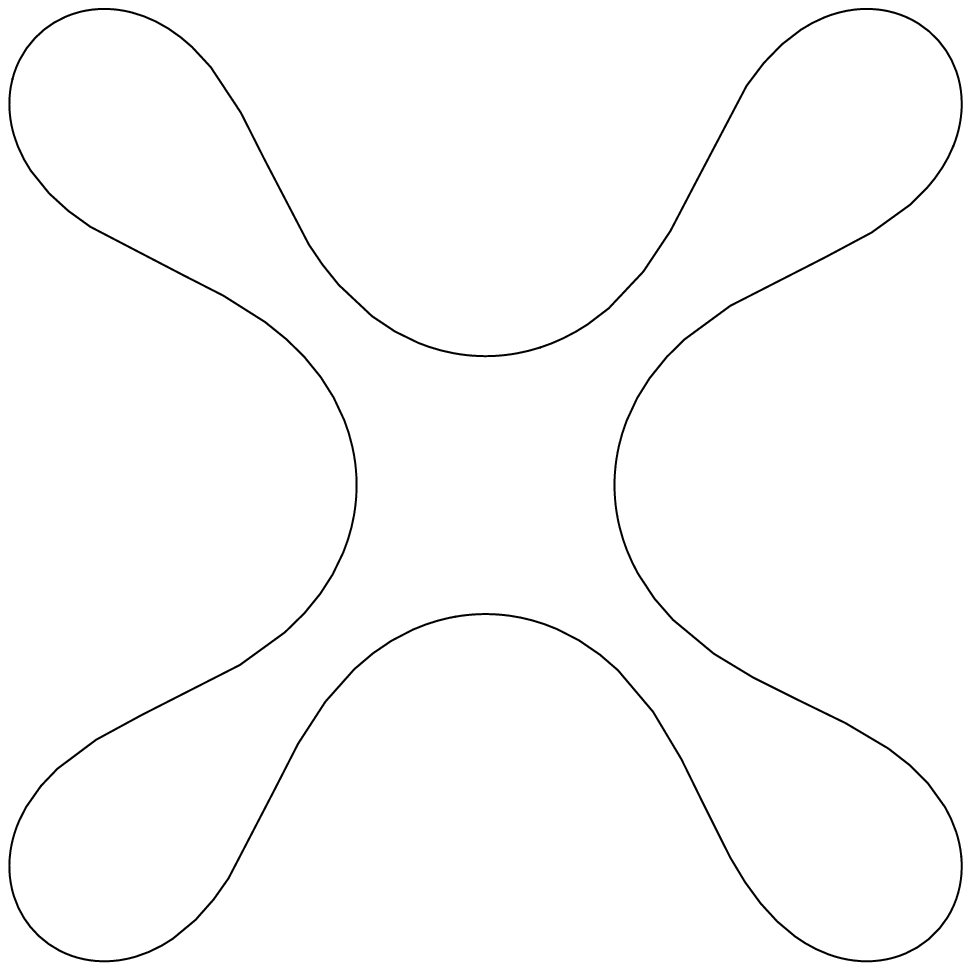}
d) \hspace*{4mm}}
\caption{Directrices of some closed cylindrical equilibrium shapes whose curvatures are solutions to equation (\ref{HEq}) of form (\ref{FK3}) with $\sigma=1$ and: \phantom{}
a) $\mu=1.908$, $E=0.146$;
b) $\mu=0$, $E=0.211$;
c) $\mu=1/3$, $E=0.407$;
d) $\mu=1/3$, $E=0.563$. }
\end{figure}

\begin{figure}[h]
\centerline{
\includegraphics[height=3.0cm]{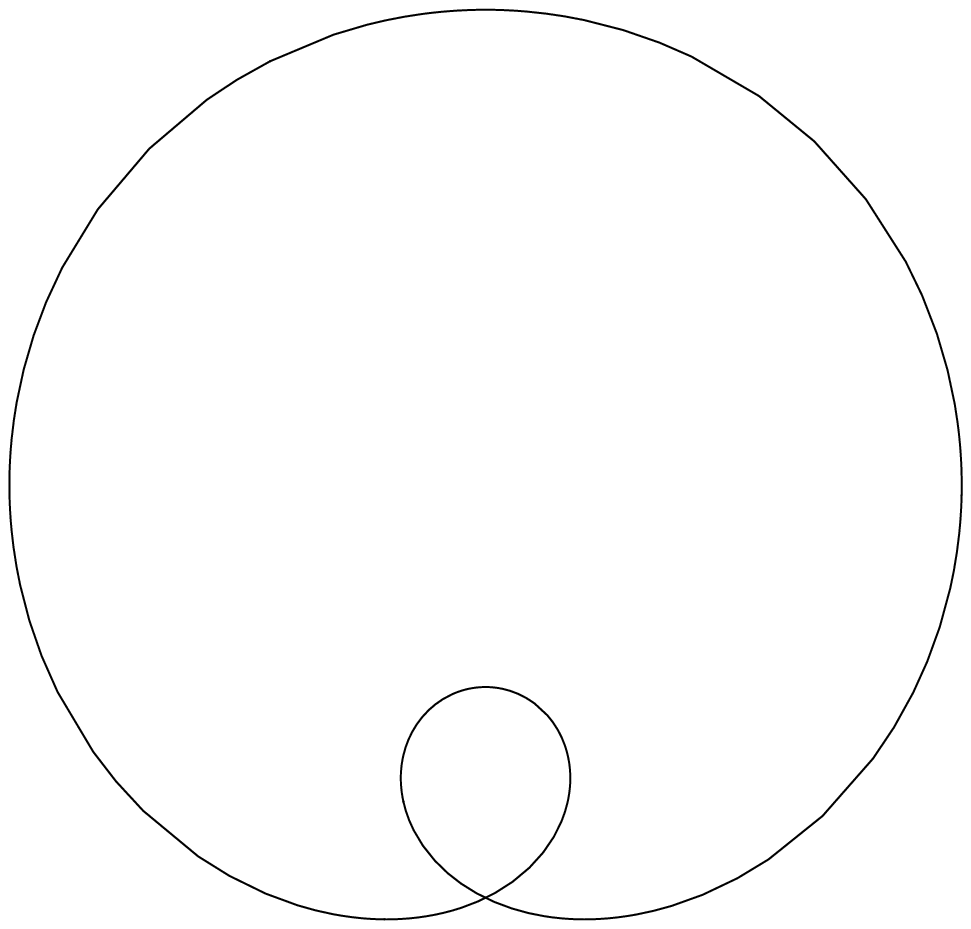}
 \hspace*{9mm}
\includegraphics[height=3.0cm]{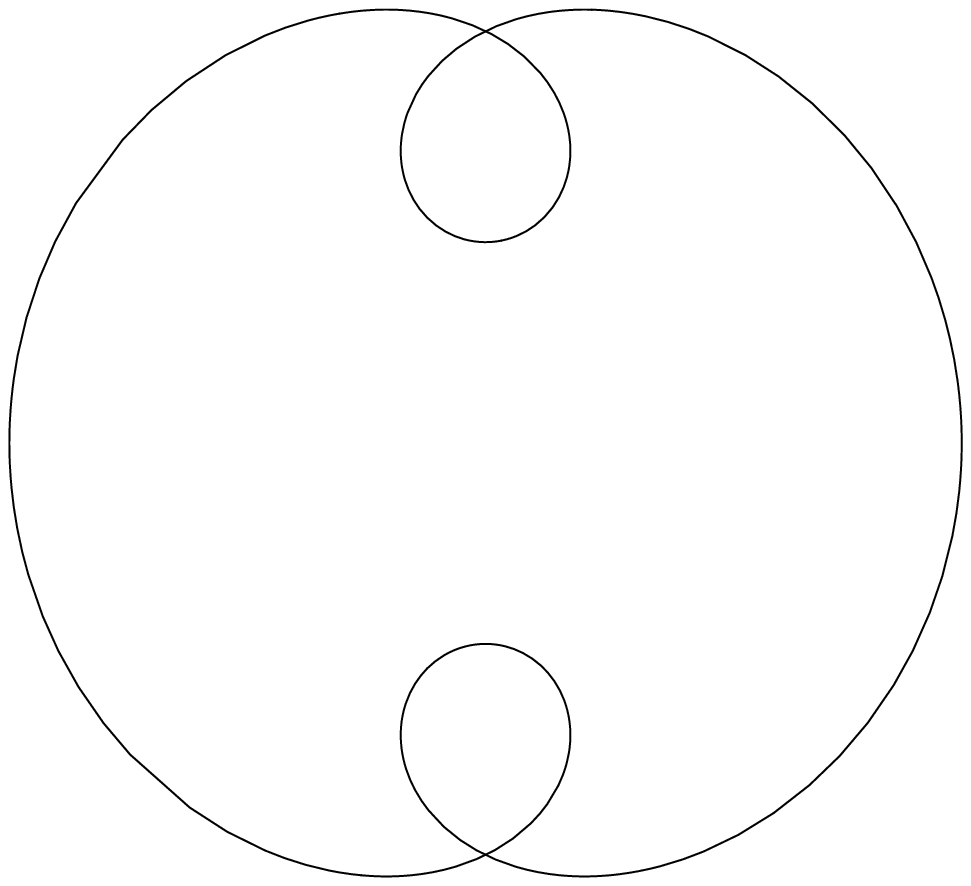}
 \hspace*{13mm}
\includegraphics[height=3.0cm]{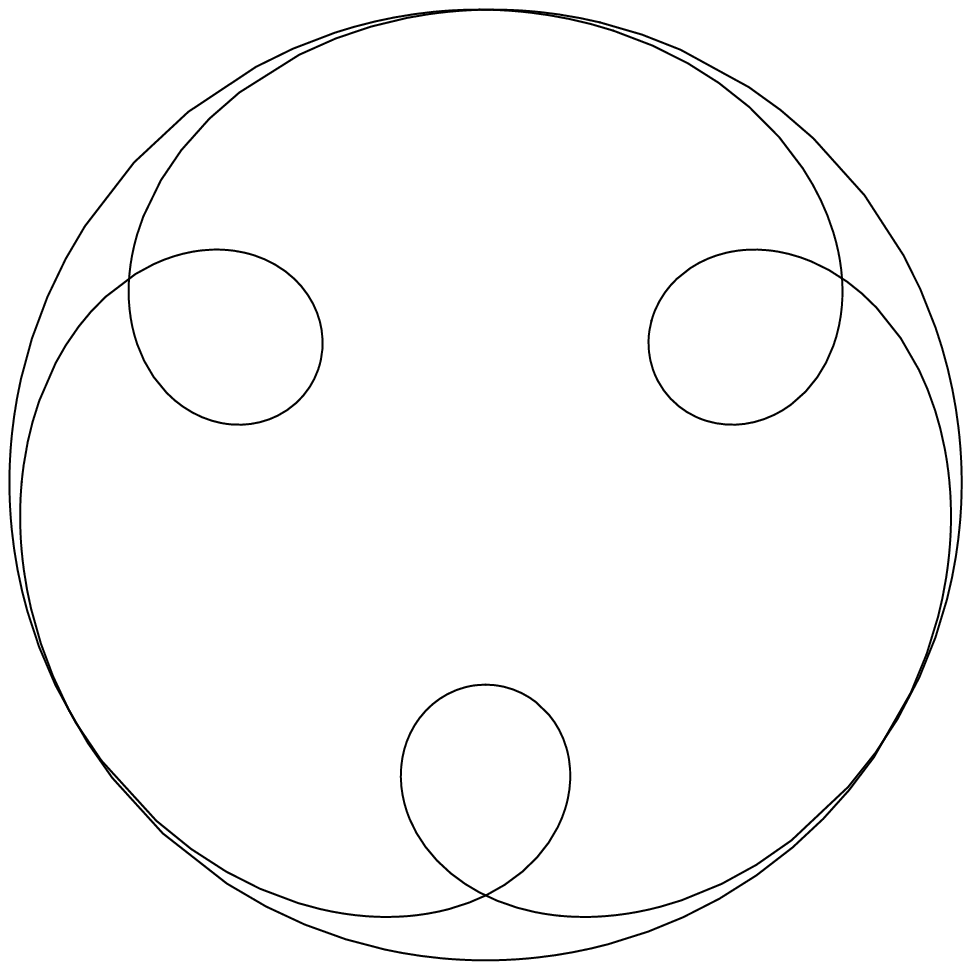}
\hspace*{0mm}
}
\centerline{
\hspace*{7mm}
\includegraphics[height=3.0cm]{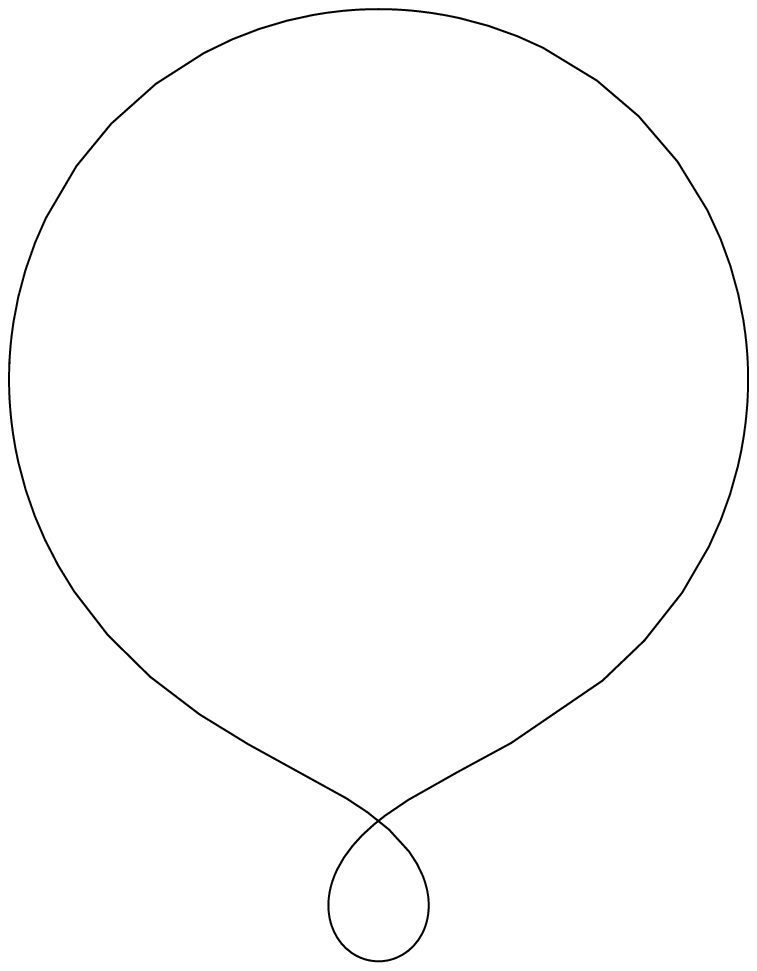}
a) \hspace*{2mm}
\includegraphics[height=3.0cm]{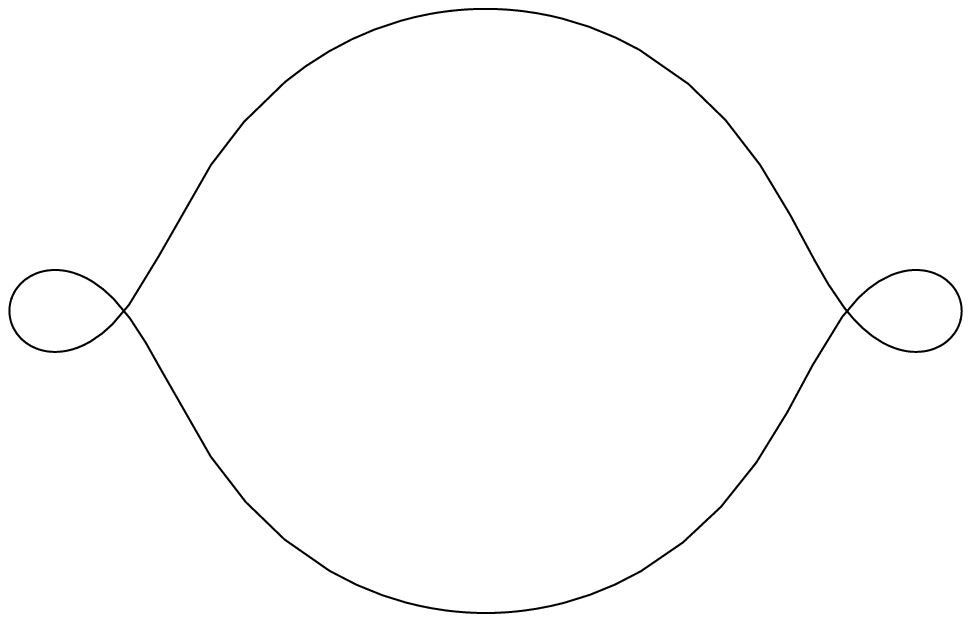}
b) \hspace*{0mm}
\includegraphics[height=3.0cm]{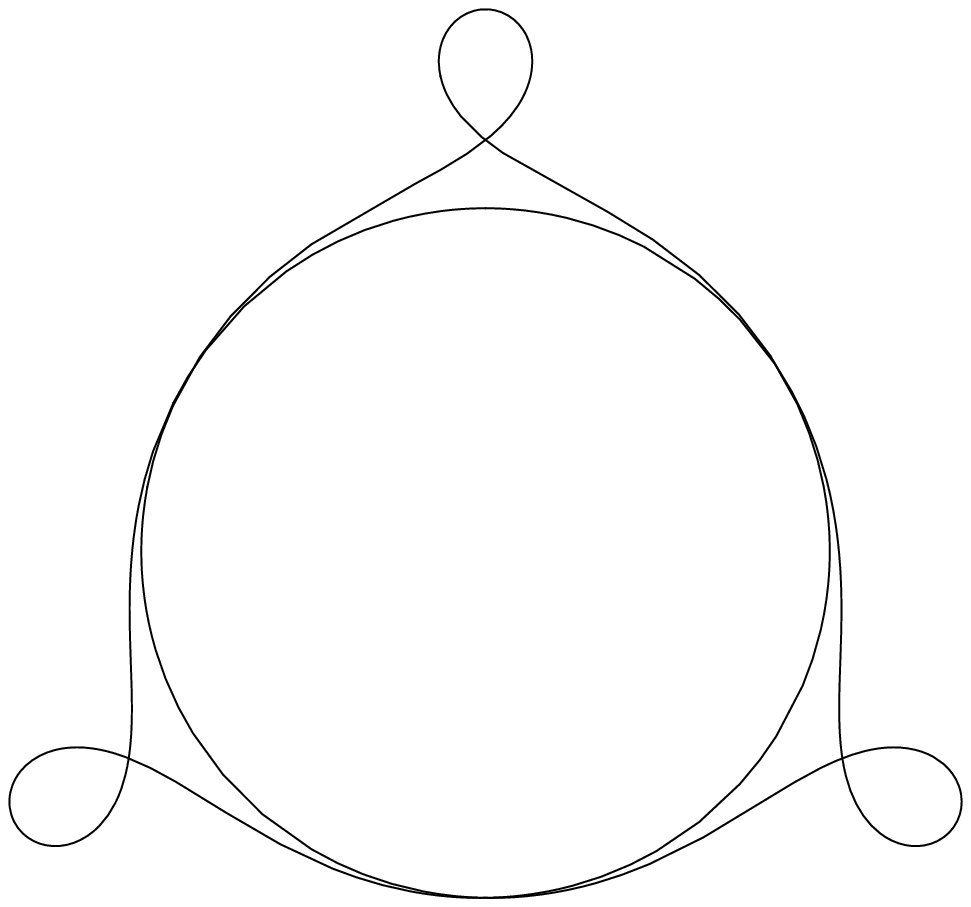}
c)
}
\caption{Directrices of some closed cylindrical equilibrium shapes whose curvatures are solutions to equation (\ref{HEq}) of forms (\ref{FK1}) (top) and (\ref{FK2}) (bottom) with $\mu=3$, $\sigma=1$ and: a) $E=0.0850056$;
b) $E=0.0849733$; c) $E=0.0850046$.
}
\end{figure}
\begin{figure}[ht]
\centerline{
\includegraphics[height=4.5cm]{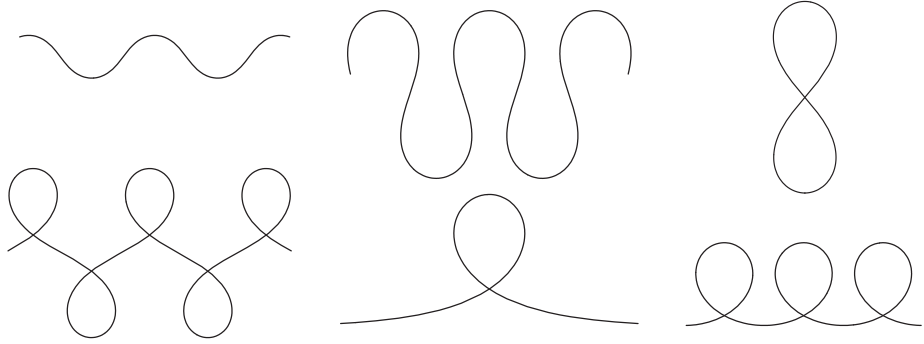}
}
\caption{Several typical equilibrium shapes of Euler's elastica related
to periodic and non-periodic solutions of equation (\ref{HEq}) with $\sigma=0$ of form (\ref{C1k}) and (\ref{C2}).
}
\end{figure}

\section{Closure conditions}

Hereafter we are interested in directrices $\Gamma $ that close up smoothly meaning that there exists a value $L$ of the arclength $s$ such that $\mathbf{x} ( 0 ) = \mathbf{x} ( L )$ and $\mathbf{t} ( 0 ) = \mathbf{t} ( L )$. The later property of such a smooth closed directrix $\Gamma $ and the definition of the tangent vector imply that there exists an integer $m$ such that  $\varphi ( L ) = \varphi ( 0 ) + 2 m \pi$ where $\varphi (s)$ is the corresponding slope angle. Since throughout this paper we always choose $\varphi ( 0 )=0$ this means that the condition $\mathbf{t} ( 0 ) = \mathbf{t} ( L )$ implies $\varphi ( L ) =  2 m \pi$.

Under the above assumptions, expressions (\ref{Sol1}) show that
\begin{equation*}
\fl
\mathbf{x}\left( 0\right) =\left( \frac{2}{\sigma }\left. \frac{{\mathrm{d}}\kappa \left( s\right) }{{\mathrm{d%
}}s}\right| _{s=0},-\frac{1}{\sigma }( \kappa\,^{2} (0) -\mu ) \right), \quad
\mathbf{x}\left( L\right) =\left( 
\frac{2}{\sigma }\left. \frac{{\mathrm{d}}\kappa \left( s\right) }{{\mathrm{d%
}}s}\right| _{s=L},-\frac{1}{\sigma }( \kappa\,^{2} ( L) -\mu
) \right). 
\end{equation*}
Consequently, the equality $\mathbf{x} ( 0 ) = \mathbf{x} ( L )$ yields \begin{equation*}
\left. \frac{{\mathrm{d}}\kappa \left( s\right) }{{\mathrm{d%
}}s}\right| _{s=L} =\left. \frac{{\mathrm{d}}\kappa ( s ) }{{\mathrm{d}}s}\right|_{s=0}, \qquad  \kappa ( L) = \pm \kappa ( 0),  
\end{equation*}
which, on account of relation (\ref{CCCL}), implies that $L $ is a period of the curvature $\kappa ( s )$, that is $L = n T $ where $n$ is a positive integer and $T$ is the least period of the function $\kappa ( s )$. Since $\varphi (n T) =  n \varphi ( T)$, as follows by formula (\ref{angle}), then $2 m \pi = \varphi (L)= \varphi (n T)=  n \varphi ( T)$ and hence
\begin{equation}
\varphi\left( T \right) = \frac {2 m \pi} {n} \, \cdot  \label{closure}
\end{equation}
Thus, in the cases when $\sigma \neq 0$, relation (\ref{closure}) is found to be a necessary condition for a directrix $\Gamma $ to close up smoothly. Apparently, it is a sufficient condition as well.

Straightforward computations lead to the following explicit expressions
\begin{equation*}
\varphi_1\left( T_1 \right) = \frac{4\left( A\beta -B\alpha
\right)}{u\left( A-B\right) } \mathrm{K}\left( k\right)
+2\frac{\left( A+B\right) \left( \alpha -\beta \right) }{u\left( A-B\right) }%
\Pi \left( -\frac{\left( A-B\right) ^{2}}{4AB},k\right),
\end{equation*}
\begin{equation*}
\varphi_3\left( T_2 \right) = \frac{2 \delta }{u}\mathrm{K}\left( k\right)+
2\frac{\alpha -\delta }{u}\Pi \left( \frac{%
\alpha -\beta }{\delta -\beta },k\right),
\end{equation*}
\begin{equation*}
\varphi_4\left( T_2 \right) = \frac{2 \beta }{u}\mathrm{K}\left( k\right) +
 2 \frac{\gamma -\beta }{u}\Pi \left( \frac{%
\gamma -\delta }{\beta -\delta },k\right),
\end{equation*}
for the angles of form (\ref {Ang33}), (\ref {Angle1}) and (\ref
{Angle2}), respectively. These expressions and the closure condition (\ref{closure}) allow to determine whether a curve of curvature (\ref {FK3}), (\ref {FK1}) or (\ref {FK2}) closes up smoothly or not.

In the cases when $\sigma=0$, relations (\ref{cs}) and (\ref{xz0}) show that
\begin{equation*}
\left. \frac{{\mathrm{d}}\kappa \left( s\right) }{{\mathrm{d%
}}s}\right| _{s=L}=0, \qquad
\kappa \left(L \right ) = \kappa ( 0), \qquad
x\left( L\right)-x\left( 0\right)=0
\end{equation*}
are the necessary and sufficient conditions for the respective directrices $\Gamma $ to close up smoothly. Let us recall that relations (\ref{cs})--(\ref{xz0}) were derived assuming $\varphi={\mathrm{d}}\kappa /{\mathrm{d}}s=0$ at $s=0$. The aforementioned conditions mean first that $L=nT$ where $n$ is a positive integer and $T$ is the least period of the function $\kappa ( s )$ and, consequently, that  
\begin{equation}
\mu T =\int_{0}^{T} \kappa \, ^{2}\left(s\right){\mathrm{d}}s,
\label{xz00}
\end{equation}
in view of the first of relations (\ref{xz0}) and due to the fact that
$\kappa(s)$ is a periodic function.

Now, using formulas (\ref{IntC1}) and (\ref{IntC2_1}), one can easily see that
\begin{equation*}
\int_{0}^{T_1}\kappa\,_1 ^{2}\left( s\right)\mathrm{d}s=
8 \sqrt {\alpha ^{2}+\eta ^{2}} \mathrm{E} \left(k\right)- \eta ^{2} T_1,
\qquad
\int_{0}^{T_2} \kappa \,_3 ^{2}\left( s\right)\mathrm{d}s=
-4 \alpha \mathrm{E} (\tilde{k})
\end{equation*}
and then, expressing the coefficient $\mu$ from formulas (\ref{mu1}) or (\ref{mu2}), to rewrite the closure condition (\ref{xz00}) for the curves of curvatures given by formulas (\ref{C1k}) or (\ref{C2}) in the form
\begin{equation*}
2 \mathrm{E} (k ) -  \mathrm{K} (k ) = 0
\label{CC01}
\end{equation*}
or
\begin{equation*}
2 \mathrm{E} (\tilde{k} ) - ( 2 - \tilde{k}^2) \mathrm{K} (\tilde{k} ) = 0,
\label{CC02}
\end{equation*}
respectively.

An interesting property of the curves of curvatures $\kappa_3 (s)$ and $\kappa_4 (s)$ is observed in the cases in which $\sigma \neq 0$ (see figure~3). Namely, if one of these curves closes up smoothly, then so does the other one. To prove this let us first note that the solutions $\kappa_3 (s)$ and $\kappa_4 (s)$ correspond to the case (II) when the polynomial $P(\kappa)$ has four real roots. Without loss of generality, these roots can be written in the form
\begin{equation}
\hspace*{-1.7cm}
\alpha = -3q-v-2w,\quad \beta = q-v-2w,\quad \gamma = q-v+2w,
\quad \delta = q+3v+2w,
\label{pvq}
\end{equation}
where $q$, $v$ and $w$ are three arbitrary positive real numbers.
The main advantage of this parametrization is that it preserves the order of the roots of the polynomial $P(\kappa)$, i.e., $\alpha<\beta<\gamma<\delta$ for any choice of the parameters $q$, $v$ and $w$, which allows one to deal freely with them.
Using these parameters, it is easy to find that
\begin{equation*}
\frac{\partial \psi}{\partial q} =
\frac{\partial \psi}{\partial v} =
\frac{\partial \psi}{\partial w} =0, \qquad
\psi=\varphi _{4}(T_2)-\varphi _{3}(T_2),
\end{equation*}
meaning that $\psi=\mathrm{const}$. This constant can be determined by evaluating the function $\psi$ for any values of the parameters $q$, $v$ and $w$, say $q=v=w$, which gives
\begin{equation}
\varphi _{4}(T_2)-\varphi _{3}(T_2)=4\pi.
\label{4Pi}
\end{equation}
This relation and the closure condition (\ref{closure}) do imply that the foregoing two curves close up simultaneously.

\section{Self-intersection}

In what concerns the vesicle shapes, of special interest are
solutions to equation (\ref{HEq})
with $\sigma \neq 0$
that give rise to closed non-self-intersecting (simple) curves. A sufficient condition for
such a closed curve to be simple is
$\mu < 0$, which is discussed in \cite{CCG}. It is also mentioned therein that the closed curves satisfying condition (\ref{closure}) with $m \neq \pm 1$ or $n= 1$ are necessarily self-intersecting. In this Section, the case $\mu>0$ is considered and several new sufficient conditions are presented for a closed curve
of the foregoing type
meeting a closure condition of form (\ref{closure}) with $m = \pm 1$ and $n\geq 2$ to be simple or not.

It is convenient to treat the problem of self-intersecting in terms of the magnitude $r(s)$ of the position vector $\mathbf{x}(s)$ and the angle $\theta (s)$ between the position vectors $\mathbf{x}(0)$ and $\mathbf{x}(s)$.
Assuming that the angle $\theta (s)$ is positive when measured counterclockwise from the vector $\mathbf{x}(0)$ to the vector $\mathbf{x}(s)$ and negative otherwise and taking into account expressions (\ref{Sol1}) we obtain the relations
\begin{equation}
x\left( s\right)  = - \mathrm{sgn} ( z\left( 0\right) ) r\left( s\right)
\sin \theta \left( s\right), \quad
z\left( s\right)  = \mathrm{sgn} ( z\left( 0\right) ) r\left( s\right)
\cos \theta \left( s\right),
\label{polar_z}
\end{equation}
\begin{equation}
\frac {\mathrm{d} \theta(s)} {\mathrm{d} s}=
\frac{\kappa\,^{2}\left( s\right) -\mu }{\sigma r\,^2(s)
}\,\cdot \label{dtheta}
\end{equation}

The following observation is crucial for the rest of the present study. 
\smallskip

\noindent \textbf{Lemma~1} \textit{Let $\Gamma $ be a smooth closed curve whose curvature $\kappa(s)$ is a solution of equation (\ref{HEq}) with $\sigma \neq 0$ of form  (\ref{FK3}), (\ref{FK1}) or (\ref{FK2}). Let $T$ be the least period of the function $\kappa(s)$ and let the corresponding slope angle meets a closure condition of form (\ref{closure}) with $m = \pm 1$ and $n\geq 2$,
i.e., $\varphi\left( T \right) = \pm 2 \pi/n$. Then, the curve $\Gamma$ is self-intersecting if and only if there exists $s_0 \in (0,T/2)$ such that $\theta(s_0) = \theta(0)$ or $\theta(s_0) = \theta(T/2)$}.\smallskip

\noindent \textbf{Proof.}
First, let us recall that each such curve has $n$ axis of symmetry
(cf. \cite{ACCG, CCG}), any position vector $\mathbf{x}(i\,T/2)$, $i=0,1,\ldots,2n$ being along one of them.
Consequently, any point of the curve whose position vector is collinear with one of the foregoing vectors lies on an axis of symmetry of the curve $\Gamma $.
Note also that on account of formulas (\ref {Ang33}), (\ref {Angle1}) or (\ref
{Angle2}) the closer condition may be written in the form $\varphi\left( T/2 \right) = \pm \pi/n$. It should be mentioned as well that formulas (\ref{FK3}), (\ref{FK1}) and (\ref{FK2}) imply ${\mathrm{d}}\kappa /{\mathrm{d}}s=0$ at $s=i\,T/2$.

Let the curve $\Gamma $ be such that
$\theta(s_0) = \theta(0)$ or
$\theta(s_0) = \theta(T/2)$
at some $s_0 \in (0,T/2)$.
Then, expressions (\ref{Sol1}), (\ref{polar_z}) and the closer condition imply $\theta\left( T/2 \right) = \pm (\pi/n + l \,\pi)$,
where $l$ is an integer,
meaning that the position vector $\mathbf{x}(s_0)$ is along a certain axis of symmetry of the curve $\Gamma $. Therefore this curve self-intersects since it passes through two different points lying on one and the same axis of symmetry.

Next, suppose that the curve $\Gamma $ is such that $\theta(s_0) \neq \theta(0)$ and
$\theta(s_0) \neq \theta(T/2)$ for each
$s_0 \in (0,T/2)$. Then, the same holds true in the next interval
$(T/2,T)$ since the curve $\Gamma $ is symmetric with respect to the axis
corresponding to the angle $\theta(T/2)$, that is along the vector $\mathbf{x}(T/2)$, and so on -- up to the last interval 
$\left(nT-T/2,nT \right)$. In this way, we arrive at the conclusion that the considered curve does not pass twice through any one of its axes of symmetry and therefore it is simple because it is simple in the interior of each of the foregoing intervals too as is evident
from relation (\ref{xx}). $\blacksquare $ \smallskip

\noindent \textbf{Theorem~3} \textit{Let $\Gamma $ be a smooth closed curve, which meets the assumptions of Lemma~1. Then}:

\textit{(i) the curve $\Gamma$ is simple if $\kappa\,^{2}(s) - \mu \neq 0$ for $s \in [0,T/2]$}; 

\textit{(ii) the curve $\Gamma $ is self-intersecting if the equation $\kappa\,^{2}(s) - \mu = 0$ has exactly one solution for $s \in [0,T/2]$}.
\smallskip

\noindent \textbf{Proof.}
({\it i}) The condition $\kappa\,^{2}(s) - \mu \neq 0$ and
expression (\ref{dtheta}) imply $\mathrm{d} \theta(s) /
\mathrm{d} s \neq 0$ for each $s \in [0,T/2]$, meaning that
$\theta(s)$ is a strictly increasing or decreasing function in
this interval. Consequently, there does not exist $s_0 \in
(0,T/2)$ such that $\theta(s_0) = \theta(0)$ or $\theta(s_0) =
\theta(T/2)$, and hence, by virtue of Lemma~1, the corresponding curve $\Gamma$ is
simple.

({\it ii}) Let $s_{\mu} $ be the only value in $[0,T/2]$ such that $\kappa\,^{2}(s_{\mu}) - \mu = 0$. First, let $s_{\mu} \in (0,T/2)$. Since $\kappa(s)$ is strictly increasing in this interval, the signs of function $\kappa\,^{2}(s) - \mu$ in the
intervals $(0,s_{\mu})$ and $(s_{\mu},T/2)$ are different. Then, expression (\ref{dtheta}) implies that the function $\theta(s)$ has an extremum at $s_{\mu}$. Suppose, $\theta(s)$ has a maximum at $s_{\mu}$. Then, when $s$
increases from $0$ to $T/2$, the function $\theta(s)$ increases
from $\theta(0) = 0$ to a certain value $\theta(s_{\mu}) = \theta_{max} >0$, and after that, decreases from $\theta_{max}$
to $\theta(T/2)$. If $\theta\left( T/2 \right)$ is negative, then there exists $ s_1 \in (s_{\mu},T/2)$ such that  $\theta(s_1) = \theta(0) $ meaning that the curve $\Gamma$ is self-intersecting. If $\theta\left( T/2 \right)$ is positive, then there exists $ s_2 \in (0,s_{\mu})$ such that  $\theta(s_2) = \theta(T/2) $ meaning that the curve $\Gamma$  is self-intersecting again. The case when $\theta(s)$ has a minimum at $s_{\mu}$ is similar. Next, let $s_{\mu} = 0$ or $s_{\mu} = T/2$. Then, expressions (\ref{Sol1}) imply $\mathbf{x}(0)=\mathbf{x}(T)=\mathbf{0}$ or $\mathbf{x}(T/2)=\mathbf{x}(3 T/2)=\mathbf{0}$, respectively. On the other hand, $\mathbf{t}(0) \neq \mathbf{t}(T)$ in the first case because $\varphi(T) = 2 \pi / n$, while in the second case $\mathbf{t}(T/2) \neq \mathbf{t}(3 T/2)$ since $\varphi(T/2) = \pi / n$ and $\varphi(3 T/2) = 3 \pi / n$. Therefore, the corresponding curves $\Gamma$ is self-intersecting.  $\blacksquare $ \smallskip

\noindent \textbf{Corollary 1} \textit{ Under the assumptions of Lemma~1, the curve $\Gamma$ is self-intersecting if its curvature $\kappa(s)$ is such that the respective polynomial $P(\kappa)$ has onlyreal roots}.
\smallskip

\noindent \textbf{Proof}.
In case (I), according to Theorem~1, real roots are achieved only if $\eta = 0$ meaning that $ \gamma = \delta = - (\alpha + \beta)/2$.
Then, formula (\ref{Coeffs1}) implies $( \alpha ^{2}-\mu ) ( \beta ^{2}-\mu ) < 0$ provided that $( 3\alpha +\beta)( \alpha +3\beta)>0$ which is the necessary and sufficient conditions for periodicity of the curvature $\kappa_1(s)$ when $\eta = 0$ (see Theorem~1). Consequently, the equation $\kappa^2_1(s) - \mu=0$ has only one solution for $s \in [0,T/2]$ since $\kappa_1(0)=\alpha$ and $\kappa_1(T/2)=\beta$. Then, according to Theorem~3 ({\it ii}), the curve is self-intersecting.

In case (II), the roots can be written in the form (\ref{pvq}) and the angles
$\varphi_{3}(T)$ and $\varphi _{4}(T)$ may be thought of as functions of the parameters $q$, $v$ and $w$. 
Differentiating relation (\ref{4Pi}) with respect to the parameter $w$, one obtains
\begin{equation*}
\frac{\rd\varphi _{3}(T)}{\rd w}=\frac{\rd\varphi _{4}(T)}{\rd w}=\frac{v^{2}-q^{2}}{w\left( q+v+w\right) \sqrt{\left( q+w\right) \left( v+w\right) }}\mathrm{E}(k). \label{Dq}
\end{equation*}
The function $\mathrm{E}(k)>1$ since $0<k<1$ and therefore the above derivatives are positive when $v>q$, negative when $v<q$ and equal to zero when $v=q$.
If $v>q$, then $\varphi _{3}(T)$ and $\varphi _{4}(T)$ are increasing functions of the variable $w$ and meet the inequalities
\begin{equation*}
\varphi _{3}(T)<-2\pi ,\qquad \varphi _{4}(T)<2\pi
\end{equation*}
since
\begin{equation*}
\lim_{w\rightarrow\infty }\varphi _{3}(T)=-2\pi ,\qquad \lim_{w\rightarrow\infty }\varphi _{4}(T)=2\pi .
\end{equation*}
The first inequality implies that if the curve corresponding to the angle $\varphi _{3}(T)$ closes up, then it necessarily self-intersects since $m<-1$ in the respective closure condition (\ref{closure}). Consider now the curve,
corresponding to $\varphi _{4}(T)$. It is easy to see that
\begin{equation*}
( \gamma ^{2}-\mu ) ( \delta ^{2}-\mu ) =4[ (q+v)^{2}+4vw] [q^{2}-3v^{2}-2qv-4vw] <0
\end{equation*}
since in this case $q^{2}-3v^{2}<0$. 
Consequently, the equation $\kappa^2_4(s) - \mu=0$ has only one solution for $s \in [0,T/2]$ since $\kappa_4(0)=\gamma$ and $\kappa_4(T/2)=\delta$.
Therefore, according to Theorem~3 ({\it ii}), the curve corresponding to $\varphi _{4}(T)$ is self-intersecting. Thus, in the case $v>q$ both curves are self-intersecting.
Similar arguments hold in the case $v<q$ and lead to the same conclusion.
Finally, if $v=q$, then the functions $%
\varphi _{3}(T)$ and $\varphi _{4}(T)$ do not depend on $w$. Evaluating them at $q=w$, one obtains
\begin{equation*}
\varphi _{3}(T) = -2\pi,  \qquad \varphi _{4}(T) = 2\pi 
\end{equation*}
and hence the two curves are self-intersecting as well. $\blacksquare $

\section{Concluding remarks}

In this paper all solutions to the membrane shape equation (\ref{MShEq}) determining cylindrical surfaces are presented in analytic form in Theorems~1 and 2. Explicit analytic expressions for the corresponding slope angles in the case $\sigma \neq 0$ are also given in these Theorems. Explicit expressions for the integrals of the squared curvatures are obtained in the case $\sigma = 0$. In this way, we have completely determined in analytic form the corresponding directrices $\Gamma $ (up to a rigid motion in the plane $\IR^{2}$) through the parametric equations (\ref{Sol1}) or (\ref{xz0}) depending on whether $\sigma \neq 0$ or $\sigma = 0$.
The necessary and sufficient conditions for the foregoing cylindrical surfaces
to be closed are derived in Section 4.
One necessary and sufficient and three sufficient conditions for their directrices to be simple or not are presented in Section 5. These conditions show that simple directrices could be achieved only in case (I), i.e., when the polynomial $P(\kappa)$ has two simple real roots and a pair of complex conjugate roots.

\ack{This research is partially supported by the Bulgarian National Science
Foundation under the grant B--1531/2005 and the contract \# 23/2006
between Bulgarian and Polish Academies of Sciences. The authors are quite grateful to Professor Clementina  Mladenova from
the Institute of Mechanics -- Bulgarian Academy of Sciences for
providing an access to her computer and computer algebra system
{\it Mathematica}$^{\circledR}$ via which  some of the
calculations presented here were done.}

\end{document}